\documentclass{article}
\usepackage{spconf,amsmath,graphicx,hyperref}
\usepackage[table]{xcolor} 
\definecolor{headerblue}{RGB}{170, 200, 240}
\definecolor{headerblue2}{RGB}{220, 235, 250} 
\usepackage{threeparttable}
\definecolor{lightred}{RGB}{255,230,230}
\definecolor{lightblue}{RGB}{215,225,240}
\definecolor{lightpurple}{RGB}{222, 193, 218}
\definecolor{lightorange}{RGB}{246, 214, 194}
\definecolor{lightgreen}{RGB}{230,255,204}
\definecolor{dorange}{RGB}{245, 184, 144}
\definecolor{orangegrey}{RGB}{230, 223, 217}
\definecolor{dblue}{RGB}{200,215,235}          
\definecolor{bluegrey}{RGB}{225,225,235}
\usepackage{tcolorbox}

\usepackage{cite}
\usepackage{tabularx} 
\newcolumntype{Y}{>{\centering\arraybackslash}X} 
\newcommand{\colorsquare}[2]{%
  \raisebox{.2ex}{\fcolorbox{black}{#1}{\rule{0pt}{#2}\rule{#2}{0pt}}}%
}
\usepackage{amssymb} 
\usepackage{bm}       

\usepackage{tabularx,array} 

\title{From Hype to Insight: Rethinking Large Language Model Integration in Visual Speech Recognition}
\name{Rishabh Jain, Naomi Harte}
\address{Sigmedia Group, School of Engineering\\
    Trinity College Dublin, Ireland\\
      \texttt{\small\{rijain, nharte\}@tcd.ie}}

\begin{document}
%
\maketitle
\begin{abstract}
Advances in self-supervised encoders have improved Visual Speech Recognition (VSR). Recent approaches integrating these encoders with LLM decoders improves transcription accuracy; however, it remains unclear whether these gains stem from visual understanding or stronger language modeling. In this work, we systematically evaluate LLM decoders by freezing or selectively updating the visual encoder, scaling decoder size, comparing adaptation strategies and architectures, and varying training data across LRS2, LRS3, and their combination. Evaluation on LRS2, LRS3, and WildVSR shows that scaling and adaptation yield limited improvements, while combining datasets enhances generalization. Semantic analysis reveals that gains arise primarily from lexical rather than semantic processing. Our Llama-2-13B model trained on the combined set achieves 24.7\% WER on LRS3 and 47.0\% on WildVSR, establishing SOTA among models trained without additional supervision. Our findings indicate LLM decoders refine contextual reasoning rather than visual features, emphasizing the need for stronger visual encoders to drive meaningful progress.

\end{abstract}
\begin{keywords}
Visual Speech Recognition, Large Language Models, AV-HuBERT, LRS datasets, Llama
\end{keywords}
\section{Introduction}
\label{sec:intro}

Visual speech recognition (VSR), or lip reading, has evolved closely alongside audio-visual speech recognition (AVSR), with both tasks leveraging similar underlying architectures and differing only in their modality inputs. These architectures have shifted from early CNN–RNN models \cite{cnn_lr} to current sequence-to-sequence (S2S) transformer-based architectures \cite{conformer_av}. The introduction of self-supervised learning (SSL) methods \cite{avhubert,Raven} marked a significant turning point, particularly with AV-HuBERT \cite{avhubert}, which leverages large amounts of unlabeled audio-visual data to learn joint representations between visual lip movements and corresponding audio signals. Building on this foundation, researchers developed variants including RAVEn \cite{Raven} and BRAVEn \cite{Braven}, which explored different self-supervised learning paradigms and architectural modifications. These approaches demonstrated that massive unlabeled pretraining can establish strong performance baselines, setting the stage for subsequent decoder-focused improvements. Semi-supervised approaches gained prominence through models like Auto-AVSR \cite{autoavsr}, which demonstrated that pseudo-labeling from pretrained ASR models could achieve competitive performance when applied to thousands of hours of unlabeled data. This approach highlighted the potential value of data-scaling strategies beyond traditional supervised learning paradigms. 

Recent research has increasingly focused on integrating pretrained visual encoders with Large Language Models (LLMs) decoders \cite{MMS-llama, mmllm_lr} to leverage LLM's linguistic knowledge for resolving ambiguities that pure visual processing cannot address \cite{wang2023visionllm}. Frameworks such as VSP-LLM \cite{vspllm} and Llama-AVSR \cite{llama_avsr} reflect this trend by connecting a pretrained visual encoder to a frozen LLM through lightweight projection layers and parameter-efficient finetuning (PeFT) approaches such as Low-Rank Adaptation (LoRA) \cite{hu2022lora} or Quantized LoRA (Q-LoRA) \cite{qlora}. These approaches aim to transfer visual speech information into the LLM, while benefiting from its pre-existing linguistic context. Although word error rate (WER) improvements are consistently reported, models trained on LRS3 \cite{lrs3} exhibit tightly clustered performance (Table~\ref{tab:vsr_lrs3}). This suggests that the observed gains may be primarily driven by stronger language modeling rather than by more effective use of the visual modality to extract lip-movement features. It remains unclear whether mapping visual features through the projection layer into the LLM produces novel visual representations.

This paper disentangles whether improvements stem from new visual features in the encoder, projection-layer learning, or other aspects of the decoder setup. To this end, we conduct experiments with a fixed visual encoder, systematically varying decoder architecture, model scale, adaptation methods, and training data strategies on both in-domain and cross-domain benchmarks. This setup isolates the decoder's contribution and clarifies how LLM integration affects the use of visual speech representations. Our findings have important implications for future research efforts in VSR.

\section{Current approaches to VSR}
\label{sec:lr}

By focusing on systems trained under comparable supervision settings, we aim to assess whether recent VSR advances reflect true architectural improvements rather than gains from additional supervision or adaptation methods. To this end, we analyze prior results on the LRS3 \cite{lrs3} and WildVSR \cite{wildvsr} datasets, considering models trained with at least 433 hours of LRS3 to ensure fair comparison. We exclude specific results from models employing additional supervision techniques, such as self-training, pseudo-labeling, or language model (LM) rescoring. Low-resource finetuning models were also excluded, except BRAVEn \cite{Braven} (30h), illustrating data scaling in Section~\ref{sec:discuss}. As summarized in Table~\ref{tab:vsr_lrs3}, WERs on LRS3 are clustered tightly between 24\% and 28\%, showing that changes in decoder design or minor training variations have limited impact. The WildVSR dataset exhibits a similar pattern. Larger gains appear mainly with pretraining or when labeled data is substantially increased, suggesting that current visual encoders may be approaching their performance limits. Models in which the AV-HuBERT encoder \cite{avhubert} was paired with LLM decoders, such as VSP-LLM \cite{vspllm} and Llama-AVSR \cite{llama_avsr}, or with the Whisper decoder \cite{av-whisper}, show only modest improvements over S2S baselines. 

\begin{table}[t]
\vspace{-10pt}
\centering
\setlength{\tabcolsep}{4pt} 
\begin{threeparttable}
\footnotesize
\caption{Previously reported results on LRS3 (L3) and WildVSR (WV) datasets for VSR.}
\label{tab:vsr_lrs3}
\begin{tabularx}{\columnwidth}{|l|c|c|>{\centering\arraybackslash}X|c|c|}
\hline
\rowcolor{headerblue}
\textbf{Model} & \textbf{PT} & \textbf{FT} & \textbf{Decoder} & \textbf{L3 $\pmb{\downarrow}$} & \textbf{WV $\pmb{\downarrow}$} \\
\hline
\rowcolor{blue!0.5} AV-HuBERT\cite{avhubert} & 1759 & 433 & S2S & 28.6 & 51.7 \\
\rowcolor{blue!0.5} AVH+Whisper\cite{av-whisper} & 1759 & 433 & Whisper & 24.3 & - \\
\rowcolor{blue!0.5} Auto-AVSR\cite{autoavsr} & -- & 818 & S2S & 33.0 & - \\
\rowcolor{blue!0.5} Auto-AVSR\cite{autoavsr} & -- & 3448 & S2S & 19.1 & 38.6 \\
\rowcolor{blue!0.5} RAVEn\cite{Raven} & 1759 & 433 & S2S & 27.8 & 52.2 \\
\rowcolor{blue!0.5} BRAVEn\cite{Braven} & 1729 & 433 & S2S & 26.6 & - \\
\rowcolor{blue!0.5} BRAVEn\cite{Braven} & 3052 & 30 & S2S & 24.8 & - \\
\rowcolor{blue!0.5} BRAVEn\cite{Braven} & 2649 & 433 & S2S & 23.6 & - \\
\rowcolor{blue!0.5} VSP-LLM\cite{vspllm} & 1759 & 433 & L-7B & 26.7 & 55.1 \\
\rowcolor{blue!0.5} VSP-LLM (E)\cite{vspllm} & 1759 & 433 & L-7B & 25.4 & 51.6 \\
\rowcolor{blue!0.5} Llama-AVSR\cite{llama_avsr} & 1759 & 433 & L-8B & 26.9 & - \\
\rowcolor{blue!0.5} Llama-AVSR (E)\cite{llama_avsr} & 1759 & 433 & L-8B & 25.3 & - \\
\rowcolor{blue!0.5} Llama-AVSR (E)\cite{llama_avsr} & 1759 & 1756 & L-8B & 24.0 & - \\
\hline
\end{tabularx}
\begin{tablenotes}[flushleft]
\small
\item  \pmb{{$\downarrow$}}: lower is better; \textbf{AVH} = AV-HuBERT;  \textbf{PT} = pretraining hours (unlabeled); \textbf{FT} = finetuning hours (labeled); \textbf{E} indicates that AV-HuBERT Encoder was updated instead of being frozen or using LoRA; \textbf{S2S} = Transformer sequence-to-sequence decoder; \textbf{L-7B} = Llama2-7B \cite{llama2}; \textbf{L-8B} = Llama3.1-8B \cite{llama3}.
\end{tablenotes}

\end{threeparttable}
\end{table}

\section{Experimental Setup}
\label{sec:format}
A key question in VSR is which factors drive performance: decoder size (1B to 13B), adaptation strategies (such as LoRA and QLoRA), LLM architectures, or training data composition. To disentangle these effects, we conduct controlled experiments evaluating their impact on recognition and reasoning. Section~\ref{sec:results} presents the results, while this section outlines the datasets, metrics, and methodology.

\subsection{Datasets and Evaluation Metrics}
\label{ssec:data_eval}

We train and evaluate our models using three standard benchmarks that together cover a wide range of VSR scenarios. \textbf{LRS2} \cite{lrs2} contains 144,482 clips, with 224 hours for training and under 1 hour each for validation and test. \textbf{LRS3} \cite{lrs3} consists of 151,819 clips from TED and TEDx talks, totaling 433 hours for training and 0.9 hours for testing. \textbf{WildVSR} \cite{wildvsr} is an open‑domain test set built using the LRS3 pipeline but sourced from unconstrained YouTube videos with greater variability in speakers, recording conditions, and visual quality, making it a challenging test set. All input videos are resampled to 16 fps; mouth regions are detected via RetinaFace \cite{RF}, cropped to 96×96 grayscale patches, and augmented with random crops during training. 

Our primary evaluation metric is WER \cite{advocating_cer}; however, for certain experiments, we also report character error rate (CER) \cite{advocating_cer} and semantic metrics such as sWER \cite{swer}, Semantic Similarity \cite{semantic_sim}, BERTScore \cite{bertscore}, and METEOR \cite{meteor}, as detailed in Section~\ref{sec:multi_eval}.

\subsection{Methodology}
\label{ssec:method}

\begin{figure}[!t]
  \centering
  \includegraphics[width=\linewidth]{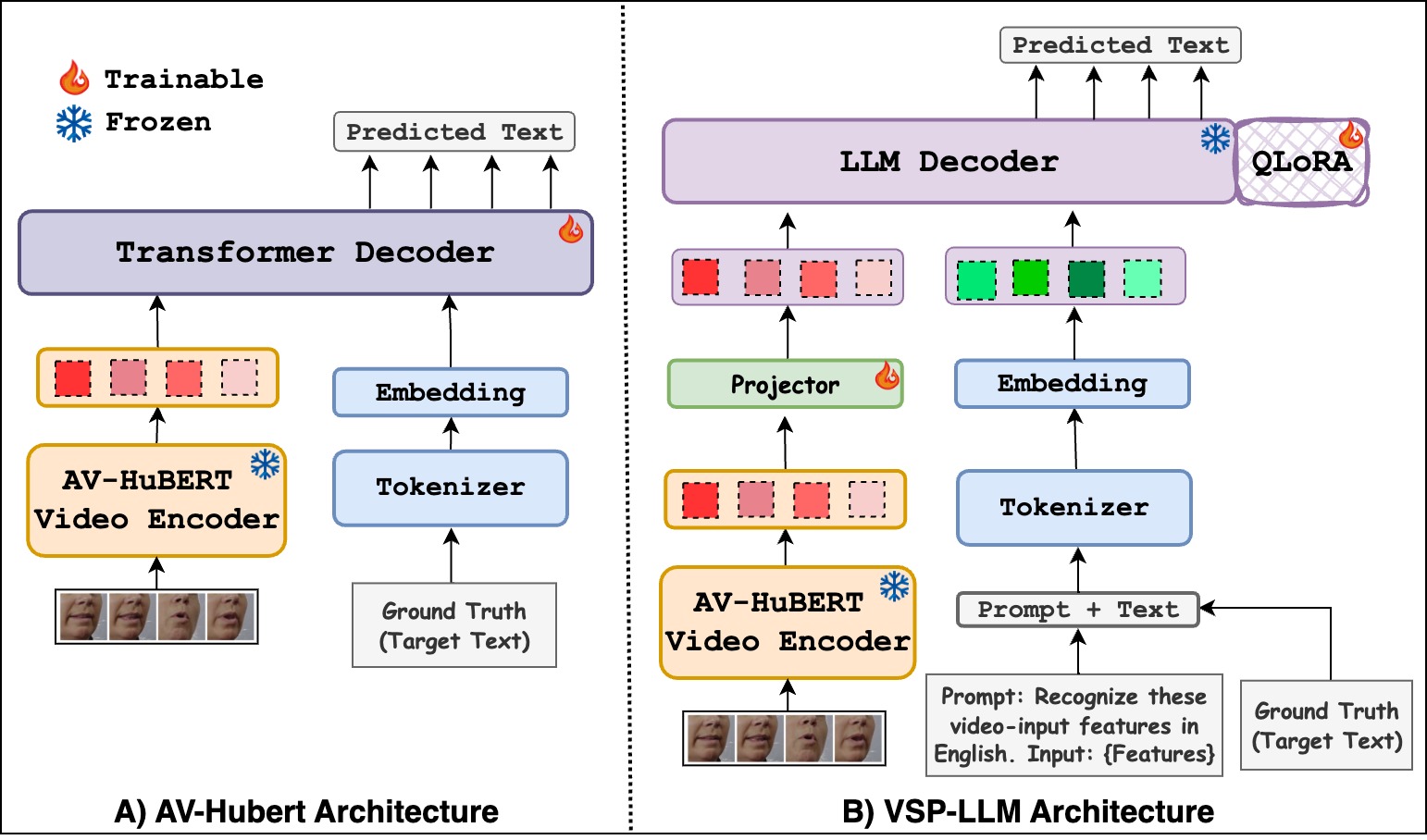}
    \caption{
    Comparison of the baseline AV-HuBERT model and the modified VSP-LLM-based architecture used in this work. 
    (A) AV-HuBERT \cite{avhubert} 
    (B) VSP-LLM \cite{vspllm}}
  \label{fig:architecture}
\end{figure}

We utilize two distinct models to investigate the impact of decoder design in VSR. The first uses the standard \textbf{AV-HuBERT}  \cite{avhubert} visual encoder with a S2S transformer decoder, serving as a strong baseline, achieving 28.6\% WER on LRS3 (see Table~\ref{tab:vsr_lrs3}). AV-HuBERT is a self-supervised framework that learns audiovisual speech representations via a hybrid ResNet-Transformer architecture. The second model builds on the \textbf{VSP-LLM} \cite{vspllm} framework, which integrates visual speech processing with LLMs to enhance contextual understanding and support multitask capabilities, achieving 26.7\% WER on LRS3 (see Table~\ref{tab:vsr_lrs3}). VSP-LLM maps visual features to the LLM input space via a visual-to-text projection layer and uses 4-bit QLoRA \cite{qlora} for efficient low-precision adaptation of LLMs. To ensure fair comparison, we exclude the deduplication step used in the original VSP-LLM. Both models use the same pretrained AV-HuBERT encoder (trained on LRS3+VoxCeleb2), ensuring that performance differences stem from decoder design rather than optimization settings. Figure~\ref{fig:architecture} illustrates the architectures of both models. In most experiments, the visual encoder is frozen and when it was unfrozen after a set number of updates, we report this explicitly. 

\section{Results and Discussion}
\label{sec:results}
\subsection{From 1B to 13B: How Decoder Size and Data Volume Drive Performance in AV-HuBERT and VSP-LLM}
\label{ssec:compare_main}
Table~\ref{tab:ft_results} shows that on the LRS2 dataset, AV-HuBERT and VSP-LLM (8B) perform identically (24.4\% WER), showing that decoder integration offers no benefit in this setting. Scaling to 13B increases in-domain error on LRS2 (28.4\% WER), suggesting overfitting, but provides modest improvement on LRS3, while performance on WildVSR remains below baseline.  When finetuning on LRS3 dataset, VSP-LLM (13B) achieves 25.7\% WER, surpassing both the 8B model and the published AV-HuBERT* baseline, and even generalizes to the challenging WildVSR testset. Training on the combined 657h corpus yields the largest gains: even the 1B decoder outperforms AV-HuBERT, and the 13B model sets new SOTA with 23.1\% WER on LRS2, 24.7\% on LRS3, and 47.0\% on WildVSR. These results highlight two key insights: (i) scaling decoder capacity yields meaningful improvements only when paired with abundant and diverse data, and (ii) LLM decoders primarily enhance contextual reasoning rather than visual representations. Notably, our 13B model trained on the combined dataset achieves the lowest WERs reported to date among models using comparable labeled resources, and excluding those that rely on additional supervision such as self-training, pseudo-labeling, or LM rescoring.

\begin{table}[t]
\vspace{-10pt}
\centering
\begin{threeparttable}
\caption{Reported WER for AV-HuBERT and VSP-LLM comparison on LRS2, LRS3, and WildVSR.}
\label{tab:ft_results}
\footnotesize
\begin{tabularx}{\columnwidth}{|X|c|c|c|}
\hline
\rowcolor{headerblue}\textbf{Model} & \textbf{LRS2 $\pmb{\downarrow}$} & \textbf{LRS3 $\pmb{\downarrow}$} & \textbf{WildVSR $\pmb{\downarrow}$} \\
\hline
\rowcolor{bluegrey}\multicolumn{4}{|l|}{\textit{Finetuned on LRS2 (224\,h)}} \\
\hline
\rowcolor{blue!0.5}AV-HuBERT & 24.4 & 38.1 & 52.9 \\
\rowcolor{blue!0.5}VSP-LLM (8B) & 24.4 & 36.4 & 56.4 \\
\rowcolor{blue!0.5}VSP-LLM (13B) & 28.4 & 36.5 & 53.6 \\
\hline
\rowcolor{bluegrey}\multicolumn{4}{|l|}{\textit{Finetuned on LRS3 (433\,h)}} \\
\hline
\rowcolor{blue!0.5}AV-HuBERT* & 38.0 & 28.7 & 51.7 \\
\rowcolor{blue!0.5}VSP-LLM (8B) & 37.9 & 27.8 & 53.8 \\
\rowcolor{blue!0.5}VSP-LLM (13B) & 38.2 & 25.7 & 49.8 \\
\hline
\rowcolor{bluegrey}\multicolumn{4}{|l|}{\textit{Finetuned on LRS2 + LRS3 (657\,h)}} \\
\hline
\rowcolor{blue!0.5}AV-HuBERT & 26.6 & 28.3 & 47.9 \\
\rowcolor{blue!0.5}VSP-LLM (1B) & 23.9 & 26.1 & 48.5 \\
\rowcolor{blue!0.5}\textbf{VSP-LLM (13B)} & \textbf{23.1} & \textbf{24.7} & \textbf{47.0} \\
\hline
\end{tabularx}
\begin{tablenotes}[flushleft]
\small
\item For all trainings, the \textbf{AV-HuBERT} encoder was updated from \textbf{22K}\pmb{$\rightarrow$}\textbf{45K}, except the baseline \textbf{AV-HuBERT*} trained with 433\,h LRS3 (as provided by the AV-HuBERT authors \cite{avhubert}); all \textbf{VSP-LLM} models are trained with QLoRA in 4-bit precision; \textbf{1B}: Llama-3.2-1B; \textbf{8B}: Llama3.1-8B; \textbf{13B}: Llama-2-13B-hf; $\pmb{\downarrow}$: lower is better.
\end{tablenotes}
\end{threeparttable}
\end{table}

\subsection{From Bits to Brain: How Quantization and Adaptation Shape Decoding}
\label{sec:typestyle}

VSR‑LLM (4-bit QLoRA, Llama‑2‑7B) and Llama‑AVSR (full-precision LoRA, Llama‑3.1‑8B) differ in adaptation and precision, yet both report nearly identical WERs: 26.7 vs. 26.9 (Table~\ref{tab:vsr_lrs3}). This raises the question of how much these decoder‑side adaptations actually influence performance. To examine this further, we designed a set of quantization experiments using Llama‑3.2‑1B, a smaller model that fits within our GPU constraints while supporting training in full precision, with LoRA, and with QLoRA. This setup, detailed in Table~\ref{tab:quant_experiment}, allows for a controlled comparison of adaptation and precision strategies. All models were finetuned for up to 30K updates on LRS3 to ensure consistency across configurations.

\begin{table}[t]
\vspace{-10pt}
\centering
\setlength{\tabcolsep}{4pt}
\begin{threeparttable}
\footnotesize
\caption{Reported WER for Quantization experiments with VSP-LLM on LRS2, LRS3, and WildVSR (WV).}
\label{tab:quant_experiment}
\begin{tabularx}{\columnwidth}{|X|c|c|c|}
\hline
\rowcolor{headerblue}
\textbf{Adaptation Approach} & \textbf{LRS2 \pmb{$ \downarrow $}}  & \textbf{LRS3 \pmb{$ \downarrow $}} & \textbf{WV \pmb{$ \downarrow $}} \\
\hline
\rowcolor{blue!0.5}QLoRA (4-bit) \cite{qlora} & 39.80 & 28.19 & 51.60 \\
\rowcolor{blue!0.5}LoRA  (16-bit) \cite{hu2022lora} & 39.35 & 28.59 & 51.93 \\
\rowcolor{blue!0.5}Full Training/ No LoRA (16-bit) & 47.33 & 37.31 & 65.43 \\
\hline
\end{tabularx}
\begin{tablenotes}[flushleft]
\small
\item \textbf{AV-HuBERT encoder} update from \textbf{22K}\pmb{$\rightarrow$}\textbf{30K} for all experiments; all models are trained with \textbf{Llama-3.2-1B} on \textbf{LRS3} dataset; the number of \textbf{bits} indicates the numerical \textbf{precision} of the model weights; \textbf{\pmb{$\downarrow$}} lower is better.
\end{tablenotes}
\end{threeparttable}
\end{table}

The results show that both 4‑bit QLoRA and 16-bit LoRA configurations achieved lower WERs than the full‑precision (no‑LoRA setup) across LRS2, LRS3, and WildVSR. In the full‑precision case, the loss was still decreasing at the end of training (at 30K updates), suggesting room for improvement with extended training. These findings suggest that quantization level and adaptation method exert only a secondary influence on accuracy, with encoder architectural constraints remaining the primary limiting factor.

\subsection{From LLM to LLM: Understanding How Decoder Architecture Affect Performance}
\label{sec:llm_arc}

To assess the impact of decoder design, we compare four similarly sized LLMs (around 13B) \cite{llama2,phi4,vicuna,qwen2.5} which are all finetuned identically on the LRS3 dataset. Although their core pipelines and training settings are similar (detailed in Table~\ref{tab:llm_architectures}), these models differ in their decoder block, positional encoding, attention variants, and pretraining data. This analysis isolates whether architectural and data-driven distinctions translate into performance gains or if the performance remains similar across designs.

\begin{table}[!t]
\vspace{-10pt}
\centering
\setlength{\tabcolsep}{4pt}
\begin{threeparttable}
\footnotesize
\caption{Impact of LLM decoder architectures and pretraining data on VSR Performance on LRS3 (L3).}
\label{tab:llm_architectures}
\begin{tabularx}{\columnwidth}{|p{1.2cm}|X|p{0.6cm}|}
\hline
\rowcolor{headerblue}
\textbf{Model} & \textbf{Architecture \& Data Details} & \textbf{L3 \pmb{$ \downarrow $}} \\
\hline
\rowcolor{blue!0.5}Llama-2-13b \cite{llama2}     & RoPE; MHA; SwiGLU; RMSNorm; 4k ctx; pretrained on 2T tokens of multilingual (web,code,dialogue); instruction-tuned & 25.7 \\
\hline
\rowcolor{blue!0.5}phi-4 \cite{phi4}          & MHA; GELU; LayerNorm; 16k ctx; English web and code; synthetic reasoning; SFT + DPO & 26.2 \\
\hline
\rowcolor{blue!0.5}vicuna-13b-v1.5 \cite{vicuna} & Extention of Llama-2 13B; RoPE; MHA; SwiGLU; RMSNorm; 4k ctx; multilingual pretraining; ShareGPT SFT (125k conversations) & 26.5 \\
\hline
\rowcolor{blue!0.5}Qwen2.5-14B \cite{qwen2.5}    & RoPE + QKV bias; GQA; SwiGLU; RMSNorm; 131k ctx; trained on 40+ lang, web, code, and structured data & 27.2 \\
\hline
\end{tabularx}
\begin{tablenotes}[flushleft]
\small
\item \textbf{T}: Trillion; \textbf{ctx}: context window; \textbf{MHA}: multi-head attention; \textbf{SFT}: supervised finetuning; \textbf{DPO}: direct preference optimization; \textbf{GQA}: grouped query attention; \textbf{RoPE}: rotary positional embeddings; \textbf{QKV}: query/key/value; models are trained till \textbf{50K} updates; Encoder updated from \textbf{22K}\pmb{$\rightarrow$}\textbf{50K}; \textbf{\pmb{$\downarrow$}}: lower is better.
\end{tablenotes}
\end{threeparttable}
\end{table}

Performance differences span within 1.5\% WER range across all tested architectures (Table~\ref{tab:llm_architectures}). Llama‑2‑13B \cite{llama2} performs best, marginally outperforming Phi‑4 \cite{phi4} design. Vicuna‑13B \cite{vicuna} and Qwen2.5‑14B \cite{qwen2.5} underperform slightly, which may be linked to their optimization on dialogue and multilingual tasks rather than general decoding. These results suggest that with a fixed encoder, architectural variations and pretraining data in the decoder only make a small difference to VSR accuracy. 
\vspace{-2.9mm}

\subsection{From Metrics to Meaning: Understanding Decoder Limits Across Lexical and Semantic Dimensions}
\label{sec:multi_eval}

To assess whether the LLM decoder introduces qualitatively different behavior compared to AV‑HuBERT, we evaluate performance using multiple complementary metrics beyond WER. While WER captures word-level accuracy, CER \cite{advocating_cer} highlights character-level errors. Semantic-WER \cite{swer} counts only substitutions that change meaning, and Semantic similarity \cite{semantic_sim} measures sentence-level closeness in embedding space. BERTScore \cite{bertscore} aligns contextual embeddings to capture semantic equivalence, and METEOR \cite{meteor} accounts for precision, recall, linguistic variations (synonyms, stemming, paraphrasing), and word order. Both AV‑HuBERT and VSP‑LLM are finetuned for 30K updates on LRS3 with the encoder frozen, ensuring that differences reflect decoder contributions rather than changes in visual representations.

As shown in Table~\ref{tab:5eval}, when evaluated on the LRS2 and LRS3 test sets, both models achieve closely matched performance on all metrics, with differences typically within 1 percentage point. VSP‑LLM occasionally achieves slightly lower WER or higher semantic score, but improvements are small and inconsistent, suggesting that the LLM decoder does not substantially alter the recognition behavior. This highlights that visual contributions come from the encoder, while decoder changes yield only minor linguistic gains.

\begin{table}[t]
\vspace{-10pt}
  \centering
  \setlength{\tabcolsep}{4pt}
  \begin{threeparttable}
\footnotesize
\caption{Extended evaluation metrics for AV-HuBERT vs VSP-LLM on 
\textit{\colorsquare{blue!1}{2.5pt} LRS2 (L2)} and 
\textit{\colorsquare{lightblue}{2.5pt} LRS3 (L3)}.}
  \label{tab:5eval}
    \begin{tabularx}{\columnwidth}{|l|Y|Y|Y|Y|>{\centering\arraybackslash}m{0.9cm}|Y|}
      \hline
      \rowcolor{headerblue}
\textbf{Model} & \textbf{CER\pmb{$\downarrow$}} & \textbf{WER\pmb{$\downarrow$}} & \textbf{sWER\pmb{$\downarrow$}} & \textbf{SS\pmb{$\uparrow$}} & \textbf{BS \pmb{$\uparrow$}} & \textbf{MET\pmb{$\uparrow$}} \\
      \hline
      \rowcolor{blue!0.5} AVH (L2) & 24.7 & 38.0 & 31.2 & 0.64 & 0.93 & 0.63 \\
      \rowcolor{blue!0.5} VLM (L2) & 26.2 & 38.3 & 31.8 & 0.64 & 0.93 & 0.61 \\
      \hline
      \rowcolor{lightblue} AVH (L3) & 18.7 & 28.7 & 22.7 & 0.71 & 0.94 & 0.70 \\
      \rowcolor{lightblue} VLM (L3) & 19.4 & 28.5 & 22.7 & 0.72 & 0.95 & 0.71 \\
      \hline
    \end{tabularx}
\par\vspace{2pt}\small
\textbf{AVH} = AV-HuBERT; \textbf{VLM} = VSP-LLM; \textbf{CER} = Character Error Rate; \textbf{WER} = Word Error Rate; \textbf{sWER} = Semantic WER; \textbf{SS} = Semantic Similarity; \textbf{BS} = BERTScore; \textbf{MET} = METEOR. All VLM use \textbf{Llama-2-7B} decoder; \pmb{$\downarrow$} lower is better; \pmb{$\uparrow$} higher is better.
  \end{threeparttable}
\end{table}

\subsection{From Bottleneck to Breakthrough: Rethinking the Visual Front-End}
\label{sec:discuss}

VSR performance is increasingly constrained by the limitations of current visual encoders. Most SOTA systems rely on AV-HuBERT or its variants \cite{llama_avsr, rouditchenko24_interspeech, vspllm}, resulting in tightly clustered WERs on LRS3 and revealing a shared ceiling in visual feature extraction regardless of decoder architecture or training strategy. The BRAVEn \cite{Braven} scaling experiment (Table~\ref{tab:vsr_lrs3}) shows that when total training duration remains the same (pretraining + finetuning), increasing labeled data 14-fold from 30 to 433 hours yields only a modest WER improvement (24.8\% to 23.6\%), highlighting the encoder as the primary bottleneck. These constraints, as revealed by our findings, highlight promising directions for advancing the VSR field. One promising approach is to develop large-scale self-supervised visual encoders trained on tens of thousands of hours of unlabeled video to learn richer lip movement representations and overcome current performance ceilings. The shift toward Vision Transformers (ViTs) \cite{vit,vivit} is also promising, as they better capture global spatio-temporal patterns than CNNs and may address key limitations in visual-phonetic mapping \cite{vallr}. Additionally, Auto-AVSR’s \cite{autoavsr} encoder, trained on 3,448 hours of pseudo-labeled data, can serve as a feature extractor, offering supervised visual representations when paired with LLM-based decoders. These findings have significant implications, indicating that further investment in decoder optimization may yield diminishing returns when performance is primarily constrained by encoder architecture. As part of future work, we aim to explore some of these directions to develop more effective VSR systems.

\section{Conclusion}
\label{sec:conclusion}

Our analysis of LLM integration in VSR shows that mapping visual features through projection layers into LLMs yields only small WER improvements, driven mainly by pretrained linguistic knowledge rather than learning new visual representations. While these gains are measurable, they do not represent the breakthroughs often suggested in recent literature. The near-universal reliance on AV-HuBERT encoders has shaped a research landscape where, despite variations in decoder design or training methods, the core visual representation learning remains largely unchanged. Our findings indicate that meaningful progress in VSR will require advances in visual encoder architectures, as decoder-focused strategies alone offer limited potential for improving VSR performance.



\section{ACKNOWLEDGMENT}
This publication emanates from research supported by Taighde Éireann – Research Ireland, Grant number 22/FFP-A/11059.


\bibliographystyle{IEEEbib}
\bibliography{refs}

\end{document}